\documentclass [12pt]{article}
\usepackage{graphicx,amssymb,amsmath,bm}
\usepackage[symbol*]{footmisc}
\usepackage[cp1251]{inputenc}
\usepackage[english,russian,ukrainian]{babel}
\usepackage{booktabs}

\makeatletter
\def\@biblabel#1{#1.\hskip-0.3em}
\makeatother

\begin{document}
\def\refname{\normalsize \centering \mdseries \bf References}
\def\abstractname{Abstract}

\begin{center}
{\large \bf Solution of Faddeev integral equations in configuration
space using the hyperspherical harmonics expansion method}
\end{center}

\begin{center}
\bf\text{V.~I.~Kovalchuk}
\end{center}

\begin{center}
\small
\textit{Department of Physics, Taras Shevchenko National University, Kiev 01033, Ukraine}
\end{center}

\begin{abstract}
A method has been developed to solve three-particle Faddeev equations in the
configuration space making use of a series expansion in hyperspherical harmonics.
The following parameters of the bound state of triton and helium-3 nuclei have
been calculated: the binding energies, the weights of symmetric and mixed-symmetry
components of the wave function, the magnetic moments, and the charge radii.
\vskip5mm
\flushleft
PACS numbers: 21.45.-v, 27.10.+h
\end{abstract}

\bigskip
\begin{center}
\bf{1.~Introduction}
\end{center}
\smallskip

Three-nucleon problems take a special place in the nonrelativistic scattering theory,
being a key to deeper physical understanding of the structure of many-particle nucleon
systems, the processes, in which such systems are engaged, and the very nature of
strong interaction. Nowadays, there exist well developed powerful methods for the solution
of three-particle problems, the most known of which are the Faddeev method~[1-4]
and the Kohn--Hulth\'{e}n variational method, the latter using a series expansion in a
basis of correlated hyperspherical harmonics~[5-7]. Benchmark calculations testify
that both approaches -- the variational and Faddeev ones -- yield very close results,
when describing experiments on $Nd$-scattering~[8,9]. The solution of a system of
coupled two-dimensional integral equations, which is required in the Faddeev method,
is not a very difficult problem for modern computers used for the description of
three-nucleon systems. However, this procedure may turn out rather resource-consuming,
when changing to systems with a nucleon number of 4 and more~\cite{10}. In recent years,
analogous problems have been attracting steady attention of theorists (see, e.g.,
Refs.~[11-13] and the references therein). It is worth noting that, in works by
Dzhibuti~[14-16], a possibility to reduce the multiplicity of integration in
the Faddeev scheme by combining the latter with the hyperspherical functions
method~[17-19] was indicated. The equivalence of both approaches was demonstrated
earlier in Ref.~\cite{20}; namely, the use of hyperspherical harmonic basis gave rise
to identical forms of final equations both in the Faddeev method and when solving a
three-particle Schr\"{o}dinger equation. The \textquotedblleft hybrid\textquotedblright\
method proposed in Refs.~[14-16] was based on the expansion of Faddeev components
in the momentum space into a series of three-particle hyperspherical eigenfunctions
mutually coupled by the Raynal--Revai unitary transformation. The bound states of triton
and $^{9}_{\Lambda}$Be hypernucleus were described~\cite{14}, and the problem of continuous
spectrum~\cite{15} (the reaction of double charge exchange of kaons at triton and helium-3
nuclei) was examined. In Refs.~\cite{21} and \cite{22}, while describing three- and
four-particle bound states, the method of hyperspherical harmonic expansion was used to
solve the Faddeev--Yakubovsky integral equations in the momentum representation. This work
aimed at studying the capabilities of hyperspherical harmonic expansion technique for the
solution of three-particle Faddeev equations in the configuration space. When calculating
the bound state parameters, we used semirealistic nucleon-nucleon potentials, which have
been used for long in similar problems  as test ones by a good many researchers.

The structure of the paper is as follows. In Sec.~2, the general formalism used when
deriving the basic equations of the method is expounded. In Sec.~3, a special case of
the problem -- a bound state of $^{3}$H and $^{3}$He nuclei -- is considered. Here,
the approximations are substantiated, and the scheme for the calculation of expansion
coefficients of the total wave function $\Psi$ is described. In Sec.~4, the calculation
results are reported for the main characteristics of three-nucleon nuclei: the binding
energy, the charge form factors, the contributions made by the symmetric and
mixed-symmetry components of $\Psi$, the magnetic moments, and the root-mean-square
charge radii. At last, in Sec.~5, a short summary of the work is made.

\bigskip
\begin{center}
\bf{2.~Formalism}
\end{center}
\smallskip

We proceed from the well-known Faddeev equations~\cite{1} written down for a system of three
strongly interacting spinless particles with identical masses $m$. Let particle~1 be scattered
by particles~2 and 3, which are in the bound state:

\begin{equation}
\begin{array}{*{20}c} {\Psi^{(1)}=\Phi+G_{0}(Z)T_{23}(Z)(\Psi^{(2)}+\Psi^{(3)})\,,}
\\ {\Psi^{(2)}=G_{0}(Z)T_{31}(Z)(\Psi^{(3)}+\Psi^{(1)})\,,}
\\ {\Psi^{(3)}=G_{0}(Z)T_{12}(Z)(\Psi^{(1)}+\Psi^{(2)})\,,}
\end{array}
\label{eq1}%
\end{equation}
where $\Psi^{(1,2,3)}$ are the one-particle wave functions, the sum of which is the total wave
function of the system,
\begin{equation}
\Psi=\Psi^{(1)}+\Psi^{(2)}+\Psi^{(3)},
\label{eq2}%
\end{equation}
$\Phi$ is the asymptotic wave function that describes the infinite motion of particle~1 with
respect to the coupled 23-pair, $G_{0}(Z)\!=\!(Z\!-\!H_{0})^{-1}$ is Green's operator,
${Z\!=\!E\pm i0}$, $E$ is the total energy of the system $1\!+\!(23)$, $H_{0}$ is the
operator of kinetic energy, and $T_{ij}$ are two-particle transition operators connected with
the pair potentials $V_{ij}$ ($ij\!=\!12,23,31$) by means of the relations:
\begin{equation}
T_{ij}(Z)=V_{ij}+V_{ij}G_{0}(Z)T_{ij}(Z)\,.
\label{eq3}%
\end{equation}
Substituting Eq.~(\ref{eq3}) into Eq.~(\ref{eq1}) and summing up the equations obtained, we
arrive at the formulas~\cite{23}:
\begin{equation}
\Psi=(1-G_{0}(Z)V_{23})\Phi+G_{0}(Z)U\Psi,\quad
U=V_{12}+V_{23}+V_{31}\,.
\label{eq4}%
\end{equation}

One can easy to show uniqueness of the solution of this equation. If you execute several
consecutive iterations for the $\Psi$, we can see that the $k$-th iteration, $\Psi_{k}$,
will not contain terms with disconnected parts except the $(G_{0}(Z)V_{23})^{k+1}\Phi$
but this term vanishes, in turn, on the $(k\!+\!1)$-th iteration, etc. Thus the whole
infinite series for $\Psi$ will contain only terms with compact kernels. Therefore,
the Eq.~(\ref{eq4}) has a unique solution.

Let us rewrite Eq.~(\ref{eq4}) as follows:
\begin{equation}
\Psi-\Phi=G_{0}(Z)U(\Psi-\Phi)+G_{0}(Z)(V_{12}+V_{31})\Phi,
\label{eq5}%
\end{equation}
and expand the difference $\Psi\!-\!\Phi$ into a series of hyperspherical harmonics (see Appendix A):
\begin{equation}
\Psi-\Phi=\sum\limits_{Kn}{\psi_{Kn}(\rho)u_{Kn}(\Omega)}.
\label{eq6}%
\end{equation}
Substituting series expansion (\ref{eq6}) into Eq.~(\ref{eq5}) and using the conditions of
orthonormalization and completeness for the eigenfunctions of the radial part of operator
$H_{0}$, as well as the condition of orthonormalization for hyperspherical harmonics, we
obtain the following system of one-dimensional integral equations for expansion coefficients
(see Appendix~B):
\begin{equation}
\psi_{Kn}(\rho)=f_{Kn}(\rho)+\lambda\sum\limits_{K^{\prime}n^{\prime}}%
{{R}_{KK^{\prime}}^{nn^{\prime}}(\rho,\bar{\rho})\,\psi_{K^{\prime}n^{\prime}%
}}(\bar{\rho}),
\label{eq7}%
\end{equation}
where
\begin{equation}
f_{Kn}(\rho)=\int{d\bar{\rho}\,}Q_{K}(\rho,\bar{\rho})\,
W_{Kn}(\bar{\rho})\,,
\label{eq8}%
\end{equation}
and
\begin{equation}
{R}_{KK^{\prime}}^{nn^{\prime}}(\rho,\bar{\rho})=\int{d\bar{\rho}\,}Q_{K}%
(\rho,\bar{\rho})\,U_{KK^{\prime}}^{nn^{\prime}}(\bar{\rho})\,
\label{eq9}%
\end{equation}
is the integral operator. The constant $\lambda$ in Eq.~(\ref{eq7}) includes numerical
coefficients and the nucleon mass $m$. The functions $Q$, $W,$ and $U$ in the integrands in
Eqs.~(\ref{eq8}) and (\ref{eq9}) look like%
\begin{eqnarray}
{Q_{K}(\rho,\bar{\rho})}&=&-(\bar{\rho}^{\,3}/\rho^{2})\Bigl\{I_{2}(\rho\,
\xi_{K}(\rho))K_{2}(\bar{\rho}\,\xi_{K}(\rho))\Theta(\bar{\rho}-\rho)\nonumber\\
&&+I_{2}(\bar{\rho}\,\xi_{K}(\rho))K_{2}(\rho\,\xi_{K}(\rho))\Theta(\rho
-\bar{\rho})\Bigr\},
\label{eq10}
\end{eqnarray}
\begin{equation}
W_{Kn}(\bar{\rho})=\int{d\Omega\,u_{Kn}^{\ast}(\Omega)(V_{12}(\bar{\rho
},\Omega)+V_{31}(\bar{\rho},\Omega))\,\Phi(\bar{\rho},\Omega)}\,,
\label{eq11}%
\end{equation}%
\begin{equation}
U_{KK^{\prime}}^{nn^{\prime}}(\bar{\rho})=\int{d\Omega\,u_{Kn}^{\ast}%
(\Omega)U(\bar{\rho},\Omega)\,u_{K^{\prime}n^{\prime}}(\Omega)}\,,
\label{eq12}%
\end{equation}
where
\begin{equation}
\xi_{K}(\rho)=\sqrt{K(K+4)/\rho^{2}-2mE},\quad E=E_{i}-\epsilon,
\label{eq13}%
\end{equation}
the quantities $I_{2}$ and $K_{2}$ in Eq.~(\ref{eq10}) are the modified Bessel functions,
whereas $E_{i}$ and $\epsilon$ in Eq.~(\ref{eq13}) are the energies of incident particle
and bound state in the three-particle system ($\epsilon\!>\!0$), respectively.
Equations~(\ref{eq7}) have the most general form. They can be solved using standard
numerical routines for an arbitrary set of hyperspherical harmonics.

The procedure used when deriving Eqs.~(\ref{eq7}) can be easily generalized to the model,
in which the wave function $\Psi$ depends on the spins and isospins of nucleons. In this
case, the difference $\Psi\!-\!\Phi$ is expanded into a series of antisymmetric basis states:
\begin{equation}
\Psi-\Phi=\sum\limits_{Kn}\psi_{Kn}(\rho){\Gamma}_{Kn}(\Omega;\sigma,\tau).
\label{eq14}%
\end{equation}
The states $\Gamma_{Kn}$ are constructed using the hyperspherical functions
$u_{Kn}^{[g]}(\Omega)$ characterized by a definite type of permutation symmetry $[g]$ --
antisymmetric, $[a]$, symmetric, $[s]$, and mixed-symmetry, $[^{\prime}]$ and
$[^{\prime\prime}]$, ones -- and the spin-isospin functions $\xi_{n}^{[\bar{g}]}(\sigma,\tau)$
with a conjugate symmetry $[\bar{g}]$ -- $[s]$, $[a]$, $[^{\prime\prime}]$, and
$[^{\prime\prime}]$, respectively (see Appendix~C). For instance, in the case of bound state
of three-nucleon system ($S\!=\!1/2$, $T\!=\!1/2$), the quantity $\Gamma_{Kn}$ looks
like~\cite{19}%
\begin{equation}
\Gamma_{Kn}=u_{Kn}^{[a]}\xi^{\lbrack s]}-u_{Kn}^{[s]}\xi^{\lbrack a]}+
u_{Kn}^{[^{\prime}]}\xi^{\lbrack^{\prime\prime}]}-u_{Kn}^{[^{\prime\prime}]}
\xi^{\lbrack^{\prime}]}.
\label{eq15}
\end{equation}

The technique used for the derivation of equations for the radial components $\psi_{Kn}(\rho)$
is similar to that presented above. Besides, the following additional orthogonality relation
for $\xi_{n}(\sigma,\tau)$ is used:
\begin{equation}
\sum\limits_{\sigma\tau}{\xi^{\lbrack g]^{\dagger}}
\xi^{\lbrack g^{\prime}]}}=\delta_{gg^{\prime}}.
\label{eq16}%
\end{equation}
In so doing, we have also to take into account, of course, that the pair potential contains
the Pauli matrices and the spin-isospin projection operators.

\bigskip
\begin{center}
\bf{3.~Bound states}
\end{center}
\smallskip

Let us consider the problem of the triton bound state making allowance for spin-isospin
degrees of freedom of its nucleons. The wave function of $^{3}$\!H can be obtained from
Eqs.~(\ref{eq5}) and (\ref{eq14}) by putting the initial condition $\Phi\!=\!0$
in them. Besides, it has also to be taken into account in Eq.~(\ref{eq13}) that $E_{i}\!=\!0$
and the total energy $E$ now acquires a sense of binding energy for $nnp$-system:
$E\!\equiv\!-E_{3}\!=\!-\epsilon$. Series (\ref{eq14}) converges rapidly~\cite{19}:
the dominant contributions to $\Psi$ are given by the partial components $\psi_{K\!=\!0,2,4}$
($n\!=\!\{0000\}$), whereas the contribution made by the partial wave with $K\!=\!6$
amounts to about 1\% of that produced by $\psi_{0}$. Without loss of generality, let us
further assume that pair interaction is central-symmetric,%
\begin{equation}
V_{ij}=\sum\limits_{s\,t}V^{(2s\!+\!1,\,2t\!+\!1)}(r_{ij})\,P_{ij}
^{(2s\!+\!1,\,2t\!+\!1)}(\sigma,\tau),
\label{eq17}
\end{equation}
where $s$ and $t$ are possible values of total spin and isospin of the $i$-th and $j$-th
nucleons, and $P_{ij}^{(2s\!+\!1,2t\!+\!1)}$ is the operator of projection onto the
spin-isospin state with the multiplicity $(2s\!+\!1,2t\!+\!1)$.

Confining expansion (\ref{eq7}) to the terms with $K$ ranging from 0 to 4 and using the
spin-isospin functions to calculate the matrix elements in the resulting system of
equations, we can ultimately write down
\begin{equation}
\begin{array}{*{20}c}
{\bigl[I+H_{00}^{(+)}\bigr]\psi_{0}=H_{02}^{(-)}\psi_{2}-H_{04}^{(+)}\psi_{4},}\\
{\bigl[2I+H_{22}^{(+)}+H_{22}\bigr]\psi_{2}=H_{20}^{(-)}\psi_{0}+H_{24}^{(-)}\psi_{4},}\\
{\bigl[I+H_{44}^{(+)}\bigr]\psi_{4}=-H_{40}^{(+)}\psi_{0}+H_{42}^{(-)}\psi_{2},} \\
\end{array}
\label{eq18}
\end{equation}
where $I$ is the unit operator,
\begin{equation}
H_{KK^{\prime}}^{(\pm)}=\frac{3\lambda}
{2}\int{d\bar{\rho}\,}Q_{K}(\rho,\bar{\rho})\int{d\Omega\,u_{Kn}^{\ast}(\Omega)}
\Big[V^{(3,1)}(\bar{\rho},\Omega)\!\pm\!V^{(1,3)}(\bar{\rho}%
,\Omega)\Big]\,u_{K^{\prime}n^{\prime}}(\Omega),
\label{eq19}
\end{equation}
\begin{equation}
H_{22}=\frac{3\lambda}
{2}\int{d\bar{\rho}\,}Q_{2}(\rho,\bar{\rho})\int{d\Omega\,u_{2n}^{\ast}(\Omega)}
\Big[V^{(3,3)}(\bar{\rho},\Omega)\!+\!V^{(1,1)}(\bar{\rho}%
,\Omega)\Big]\,u_{2n^{\prime}}(\Omega)
\label{eq20}%
\end{equation}
are integral operators, and $\lambda\!=\!16m/\pi$. The multi-indexes $n$ and $n^{\prime}$ are
{$n\!=\!n^{\prime}\!=\!\{0000\}$} in (\ref{eq19}) and {$n\!=\!n^{\prime}\!=\!\{1100\}$}
in (\ref{eq20}).

The following routine was used to solve system (\ref{eq18}). Presenting the latter in
the matrix form and zeroing the right hand sides in Eqs.~(\ref{eq18}), it is easy to
verify directly that, for the $NN$-potentials used in this work (see below), only the
first of the three resulting homogeneous equations has a nontrivial solution. Therefore,
the iterative procedure becomes determined unambiguously. First, as a first approximation,
we find $\phi_{0}$ from the equation $\phi_{0}\!=\!-H_{00}^{(+)}\phi_{0}$. The nonzero
solution of this equation (the eigenfunction $\phi_{0}$) is known~\cite{24} to exist, provided
that the matrix of integral operator $H_{00}^{(+)}$ is degenerate. By solving the equation
$\det H_{00}^{(+)}(E)\!=\!0$, we find the eigenvalue $E$, the sense of which is the binding
energy of triton. Afterwards, the function $\psi_{0}$ determined to an accuracy of a constant
factor is substituted into the second and third equations of system (\ref{eq18}) to find
$\psi_{2}$ and $\psi_{4}$. The normalization constant is determined from the condition
\begin{equation}
\int d\mathbf{x}d\mathbf{y}|\Psi|^{2}=P_{S}+P_{S^{\prime}}=1,
\label{eq21}%
\end{equation}
where
\begin{equation}
P_{S}=\int d\rho{\rho}^{5}(\psi_{0}^{2}(\rho)+\psi_{4}^{2}(\rho))
\label{eq22}%
\end{equation}
and
\begin{equation}
P_{S^{\prime}}=\int d\rho{\rho}^{5}\psi_{2}^{2}(\rho)
\label{eq23}%
\end{equation}
are the weights of symmetric and mixed-symmetry, respectively, components of the triton
wave function. The $S$- and $S^{\prime}$-components combine the states with $L\!=\!0$\
and $S\!=\!1/2$. The small $P$-component ($L\!=\!1$; $S\!=\!1/2,\,3/2$), the contribution
of which to the normalization integral is of the order of 0.1\%, was not taken into account.

The routine used to find the wave function and the binding energy, which was described above,
can also be applied to the $^{3}$\!He nucleus. In this case, the kernels of integral operators
(\ref{eq19}) and (\ref{eq20}) must be appended with the Coulomb term $2V_{C}/3$, where $V_{C}$
is the Coulomb potential. Certainly, the component with isospin $T\!=\!3/2$ is neglected in
the $^{3}$\!He wave function determined from Eqs.~(\ref{eq18}) using this routine. However, as
was shown in Ref.~\cite{25}, its contribution to normalization integral (\ref{eq21}) does
not exceed $2.5\cdot10^{-3}$ \%.

\bigskip
\begin{center}
\bf{4.~Calculation results}
\end{center}
\smallskip

The system of equations (\ref{eq18}) was solved for a triton in the cases, when the
Malfliet--Tjon~\cite{26},
\begin{equation}
\begin{array}{*{20}c}
{V_{t}(r)=(1438.72\,e^{-3.11r}-626.885\,e^{-1.55r})/r,} \\
{V_{s}(r)=(1438.72\,e^{-3.11r}-513.968\,e^{-1.55r})/r,} \\
\end{array}
\label{eq24}%
\end{equation}
Volkov~\cite{27,28},
\begin{equation}
\begin{array}{*{20}c}
{V_{t}(r)=144.86\,e^{-(r/0.82)^2}-83.34\,e^{-(r/1.6)^2},} \\
{V_{s}(r)\!=\!0.63V_{t}(r),} \\
\end{array}
\label{eq25}%
\end{equation}
and Eikemeier--Hackenbroich~\cite{29},
\begin{equation}
\begin{array}{*{20}c}
{V_{t}(r)=600\,e^{-5.5r^2}-70\,e^{-0.5r^2}-27.6\,e^{-0.38r^2},} \\
{V_{s}(r)=880\,e^{-5.4r^2}-70\,e^{-0.64r^2}-21\,e^{-0.48r^2}} \\
\end{array}
\label{eq26}%
\end{equation}
potentials are used.

The determined dependencies $\psi_{K}(\rho)$ were used to calculate the charge form
factors of $^{3}\mbox{H}$ and $^{3}\mbox{He}$ nuclei~\cite{30,31}
\begin{equation}
\begin{array}{*{20}c}
{F_{ch}(^{3}\mbox{H})=2F_{ch}^{n}F_{L}+F_{ch}^{p}F_{O},} \\
{2F_{ch}(^{3}\mbox{He})=2F_{ch}^{p}F_{L}+F_{ch}^{n}F_{O},} \\
\end{array}
\label{eq27}%
\end{equation}
where
\begin{equation}
\begin{array}{*{20}c}
{F_{L}=F_{1}-F_{2}/3,} \\
{F_{O}=F_{1}+2F_{2}/3.} \\
\end{array}
\label{eq28}%
\end{equation}

In turn, the bulk form factors $F_{1}$ and $F_{2}$ in Eq.~(\ref{eq28}) are expressed
in terms of $\psi_{K}$ as follows~\cite{32}:
\begin{equation}
{F_{1}(q)}=24\sqrt{3}\int\frac{{\rho}^{5}d\rho}{a^{2}}\,\bigl[\psi_{0}^{2}
J_{2}(a)-2\sqrt{3}\psi_{0}\psi_{4}J_{6}(a)
+\psi_{2}^{2}(J_{2}(a)+J_{6}(a))\bigr],
\label{eq29}%
\end{equation}
\begin{equation}
F_{2}(q)=72\sqrt{6}\int\frac{{\rho}^{5}d\rho}{a^{2}}\,\psi_{0}\psi_{2}%
J_{4}(a),
\label{eq30}%
\end{equation}
where $a=\sqrt{2/3}q\rho$.

Using the known formula for the charge form factor at small $q$'s,
\begin{equation}
F_{ch}(q)=1-q^{2}r_{ch}^{2}/6+...\,
\label{eq31}%
\end{equation}
and Eqs.~(\ref{eq27})--(\ref{eq30}), we can obtain~\cite{32} the following expressions
in the approximation $F_{ch}^{n}=0$:
\begin{equation}
\begin{array}{*{20}c}
{r_{ch}^{2}(^{3}\mbox{H})=(r_{ch}^{p})^{2}+r_{+}^{2},} \\
{r_{ch}^{2}(^{3}\mbox{He})=(r_{ch}^{p})^{2}+r_{-}^{2},} \\
\end{array}
\label{eq32}%
\end{equation}
where $r_{ch}^{p}=0.842~\mathrm{fm}$ is the proton charge radius~\cite{33}, and
\begin{equation}
r_{\pm}^{2}=\int\rho^{7}d\rho(\sqrt{3}\psi_{0}^{2}\pm\frac{\sqrt{6}}{4}%
\psi_{0}\psi_{2}).
\label{eq33}%
\end{equation}

The magnitudes of $^{3}\mbox{H}$ and $^{3}\mbox{He nuclear}$ magnetic moments can be
calculated by expressing them in terms of known magnetic moments of proton, $\mu_{p}$,
and neutron, $\mu_{n}$, and the weights of symmetric, $P_{S}$, and mixed-symmetry,
$P_{S^{\prime}}$, components~\cite{34}:
\begin{equation}
\mu=\frac{\mu_{p}+\mu_{n}}{2}(P_{S}+P_{S^{\prime}})-T_{3}\frac{\mu_{p}-
\mu_{n}}{2}(P_{S}-\frac{1}{3}P_{S^{\prime}}),
\label{eq34}%
\end{equation}
where $T_{3}=1/2$ for $^{3}$He and $T_{3}=-1/2$ for $^{3}$H.

In the table~1, the calculated binding energies, the weights of wave function components,
the magnetic moments, and the root-mean-square charge radii for $^{3}\mbox{H}$ and
$^{3}\mbox{He}$ nuclei are quoted.

\bigskip
\begin{center}
\small
Table~1. Properties of $^{3}\mbox{H}$ and $^{3}\mbox{He}$ nuclei calculated for various models
of $NN$-potential.
\end{center}
\smallskip
\vspace{-5mm}
\begin{table}[!h]
\small
{\begin{tabular}{@{}cccccc@{}}
\hline
& Malfliet-- & Volkov & Eikemeier-- & Argonne $\textit{v}_{18}$ + & Experiment \\
& Tjon & & Hackenbroich & Urbana IX,~\cite{24}  &  \\
\hline
$E(^3\mbox{H}$), MeV                   &    7.981 &    7.665  &    8.942  &   8.479  &    8.482~\cite{35}  \\
$P_{S}(^3\mbox{H}, K\!=\!0)$, \%       &   95.91  &   95.75   &   97.03   &          &                    \\
$P_{S}(^3\mbox{H}, K\!=\!4)$, \%       &    2.67  &    2.54   &    1.76   &          &                    \\
$P_{S'}(^3\mbox{H})$, \%               &    1.42  &    1.71   &    1.21   &   1.05   &                    \\
$\mu(^3\mbox{H}), \mbox{nucl. magn.}$  &    2.748 &    2.739  &    2.755  &          &    2.979~\cite{36}  \\
$r_{ch}(^3\mbox{H}), \mbox{fm}$        &    1.667 &    1.692  &    1.644  &          &    1.755~\cite{37}  \\
$E(^3\mbox{He}$), MeV                  &    7.241 &    6.951  &    8.165  &   7.750  &    7.719~\cite{35}  \\
$P_{S}(^3\mbox{He}, K\!=\!0)$, \%      &   95.83  &   95.89   &   96.91   &          &                    \\
$P_{S}(^3\mbox{He}, K\!=\!4)$, \%      &    2.64  &    2.24   &    1.74   &          &                    \\
$P_{S'}(^3\mbox{He})$, \%              &    1.53  &    1.87   &    1.35   &   1.24   &                    \\
$\mu(^3\mbox{He}), \mbox{nucl. magn.}$ & -- 1.865 & -- 1.855  & -- 1.871  &          & -- 2.127~\cite{36}  \\
$r_{ch}(^3\mbox{He}), \mbox{fm}$       &    1.814 &    1.777  &    1.737  &          &    1.959~\cite{37}  \\
\hline
\end{tabular}}
\end{table}

\vspace{-2mm}
\begin{center}
\bf{5.~Conclusions}
\end{center}
\smallskip

The study of scattering processes in systems composed of three strongly interacting particles
has been a subject of enhanced interest of researchers for a long time. However, only in the
first half of the 1990s, the methods were developed, which allowed high-precision calculations
of observable quantities in $3N$-reactions to be carried out. The method of Faddeev equations
and the method of hyperspherical harmonics, which belong to the most known and effective
approaches in researching $3N$-systems, are deservedly classed as such. In this work, those
two approaches have been combined together; namely, the series expansion in hyperspherical
harmonics was used to find the solutions of Faddeev integral equations in the configuration
space. The new approach takes advantage of the problem geometry directly, by representing the
solution of Faddeev equations as a series in the eigenfunctions of the angular part of
six-dimensional Laplace operator (the hyperspherical harmonics). As a result, the problem is
reduced to the solution of a system of one-dimensional integral equations valid for an
arbitrary potential. From a comparison between the results obtained and high-precision data
(see Table 1), it follows that the method proposed allows the basic characteristics of the
bound state of $^{3}\mbox{H}$ and $^{3}\mbox{He }$nuclei to be described satisfactorily for
the approximations and potentials used in this work. An advantage of the method is also the
circumstance that, unlike the works by Dzhibuti~\cite{14,15}, it does not use the
Raynal--Revai transformation for partial components of hyperspherical functions, because it
is the total wave function that is expanded into a series of hyperspherical harmonics.

\setcounter{section}{0}
\def\theequation{\Alph{section}.\arabic{equation}}
\def\thesection{\normalsize Appendix \Alph{section}:}
\setcounter{equation}{0}
\section{\normalsize{Hyperspherical harmonics}}
\hspace{\parindent}
The general relations for three-particle hyperspherical harmonics look like~\cite{16}
\begin{equation*}
u_{K}^{l_{x}l_{y}LM}(\Omega)=\sum\limits_{m_{x}m_{y}}{\bigl(l_{x}l_{y}%
m_{x}m_{y}|LM\bigr)}u_{K}^{l_{x}l_{y}m_{x}m_{y}}(\Omega),
\end{equation*}

\begin{eqnarray}
{u_{K}^{l_{x}l_{y}m_{x}m_{y}}(\Omega)}&=&N_{K}^{l_{x}l_{y}}(\cos{\theta})^{l_{x}}
(\sin{\theta})^{l_{y}}\nonumber\\
&&\times P_{q}^{l_{y}+1/2,\,l_{x}+1/2}(\cos{2\theta})\,
Y_{l_{x}m_{x}}(\mathbf{{\hat{x}}})Y_{l_{y}m_{y}}(\mathbf{{\hat{y}}}),
\label{eqA1}%
\end{eqnarray}
where
\begin{equation*}
N_{K}^{l_{x}l_{y}}=\sqrt{\frac{2q{\kern 1pt}!(K+2)(q+l_{x}+l_{y}+1)}
{\Gamma(q+l_{x}+3/2)\Gamma(q+l_{y}+3/2)}},
\end{equation*}
\begin{equation*}
q\!=\!\frac{K\!-\!l_{x}\!-\!l_{y}}{2},
\end{equation*}

\begin{equation*}
P_{q}^{\alpha,\,\beta}(z)=2^{-q}\sum\limits_{p=0}^{q}{\dbinom{q+\alpha}{p}}
{\dbinom{q+\beta}{q-p}}(z-1)^{q-p}(z+1)^{p}
\end{equation*}
is the Jacobi polynomial, ${\dbinom{a}{b}}\!=\!\dfrac{\Gamma(a+1)}{b!\,\Gamma(a-b+1)}$,
$l_{x}$ is the pair orbital moment corresponding to the Jacobi coordinate
$\mathbf{x}\!=\!\sqrt{1/2}(\mathbf{r}_{2}\!-\!\mathbf{r}_{3})$,
and $l_{y}$ is the orbital momentum of the first particle with respect to the pair center
of mass corresponding to the Jacobi coordinate
$\mathbf{y}\!=\!\sqrt{2/3}(\mathbf{r}_{1}\!-\!(\mathbf{r}_{2}\!+\!\mathbf{r}_{3})/2)$.

The notation $u_{Kn}(\Omega)\!\equiv u_{K}^{l_{x}l_{y}LM}(\Omega)$ for hyperspherical
harmonics includes the following quantities: $K$ is the hypermoment; $n$ is a multisubscript,
which includes the orbital moments $l_{x}$ and $l_{y}$, the total orbital moment $L$ of
the relative motion of all three particles, and its projection $M$; and
$\Omega\!=\!\{\Theta,\theta_{x},\phi_{x},\theta_{y},\phi_{y}\}$ is the set of five angles
in the six-dimensional space, which determine the orientation of the six-dimensional vector
$\mathbf{\rho}\!=\!\sqrt{\mathbf{{x}}^{2}\!+\!\mathbf{{y}}^{2}}$.

\setcounter{equation}{0}
\section{\normalsize{System of integral equations for expansion coefficients}}
\hspace{\parindent}
The kinetic energy operator $H_{0}$ in the hyperspherical basis
$(\rho,\Omega)$ looks like
\begin{equation}
H_{0}=T_{0}-\frac{\hbar^{2}}{2m\rho^{\,2}}\,\Delta_{\Omega}\,,\quad
T_{0}=-\frac{\hbar^{2}}{2m\rho^{\,5}}\,\frac{\partial}{\partial\rho}\rho
^{\,5}\frac{\partial}{\partial\rho}.
\label{eqB1}
\end{equation}
The eigenfunctions of operator $\Delta_{\Omega}$ are hyperspherical
harmonics (\ref{eqA1})
\begin{equation}
\Delta_{\Omega}u_{Kn}(\Omega)=-K(K+4)u_{Kn}(\Omega).
\label{eqB2}
\end{equation}
The functions $u_{Kn}(\Omega)$ are mutually orthogonal. They are normalized
in a standard way:
\begin{equation}
\int{d\Omega}\,u_{Kn}^{\ast}(\Omega)\,u_{K^{\prime}n^{\prime}}(\Omega)=
\delta_{KK^{\prime}}\,\delta_{nn^{\prime}}\,.
\label{eqB3}
\end{equation}
The eigenfunctions $\omega_{q}(\rho)$ of operator $T_{0}$ in Eq.~(\ref{eqB1})
are determined by the equation~\cite{38}
\begin{equation}
T_{0}\,\omega_{q}(\rho)=\frac{q^{\,2}}{2m}\,\omega_{q}(\rho)\,,\quad
\omega_{q}(\rho)=\frac{\sqrt{q}}{\rho^{\,2}}\,J_{2}(q\rho).
\label{eqB4}%
\end{equation}
They satisfy the orthonormalization,
\begin{equation}
\int\limits_{0}^{\infty}d\rho\,\rho^{\,5}\,\omega_{q}^{\ast}(\rho)\,
\omega_{q^{\prime}}(\rho)=\delta(q-q^{\prime}),
\label{eqB5}%
\end{equation}
and completeness,
\begin{equation}
\int\limits_{0}^{\infty}dq\,\omega_{q}^{\ast}(\rho)\,
\omega_{q}(\rho^{\prime})=\frac{1}{\rho^{\,5}}\,\delta(\rho-\rho^{\prime})
\label{eqB6}%
\end{equation}
conditions.

Substituting Eq.~(\ref{eq6}) into Eq.~(\ref{eq5}), multiplying both sides
of the obtained equation by $u_{K^{\prime}n^{\prime}}^{\ast}$, integrating
over $\Omega$-variables, and taking into account Eq.~(\ref{eqB3}),
we obtain the system of integral equations for
$\psi_{K^{\prime}n^{\prime}}(\rho)$:
\begin{eqnarray}
{\psi_{K^{\prime}n^{\prime}}(\rho)}&=&\int d\Omega\,u_{K^{\prime}n^{\prime}}
^{\ast}(\Omega)G_{0}(Z)f(\rho,\Omega)\nonumber\\
&&+\int d\Omega\,u_{K^{\prime}n^{\prime}}^{*}(\Omega)
G_{0}(Z)(V_{12}+V_{31})\Phi,
\label{eqB7}
\end{eqnarray}
\begin{equation}
f(\rho,\Omega)=U\sum\limits_{Kn}{\psi_{Kn}(\rho)}u_{Kn}(\Omega),
\label{eqB8}%
\end{equation}
where $U$ is the sum of three pair potentials (\ref{eq4}).

Now, substitute Green's operator
\begin{equation}
G_{0}(\Omega)=\Bigl[Z-T_{0}+\frac{\hbar^{2}}{2m\rho^{2}}
\Delta_{\Omega}\Bigr]^{-1}
\label{eqB9}
\end{equation}
into Eq.~(\ref{eqB7}) and expand the function $f(\rho,\Omega)$
(see Eq.~(\ref{eqB8})) into a series of hyperspherical harmonics:
\begin{equation}
f(\rho,\Omega)=\sum\limits_{K^{\prime\prime}n^{\prime\prime}}
{f_{K^{\prime\prime}n^{\prime\prime}}(\rho)}
u_{K^{\prime\prime}n^{\prime\prime}}(\Omega),
\label{eqB10}
\end{equation}
\vspace{1pt}
\begin{equation}
f_{K^{\prime\prime}n^{\prime\prime}}(\rho)=\int d\Omega\,u_{K^{\prime\prime}
n^{\prime\prime}}^{\ast}(\Omega)f(\rho,\Omega).
\label{eqB11}
\end{equation}
Then, taking Eqs.~(\ref{eqB2}) and (\ref{eqB3}) into account, the system
of equations (\ref{eqB7}) reads
\begin{equation}
\psi_{K^{\prime}n^{\prime}}(\rho)=\Bigl[Z-T_{0}-\frac{\hbar^{2}}{2m\rho^{2}}
K^{\prime}(K^{\prime}+4)\Bigr]^{-1}F(\rho),
\label{eqB12}
\end{equation}
\begin{equation}
F(\rho)=\int d\Omega\,u_{K^{\prime}n^{\prime}}^{\ast}(\Omega)
\Bigl[f(\rho,\Omega)+(V_{12}+V_{31})\Phi\Bigr].
\label{eqB13}
\end{equation}
Now, expand the function $F(\rho)$ in a series of complete system of
$T_{0}$-operator (\ref{eqB4}) eigenfunctions,
\begin{equation}
F(\rho)=\int\limits_{0}^{\infty}dq\,\omega_{q}(\rho)F_{q},
\label{eqB14}
\end{equation}
\begin{equation}
F_{q}=\int\limits_{0}^{\infty}d{\bar{\rho}}\,
{\bar{\rho}}^{\,5}\,F(\bar{\rho})\omega_{q}^{\ast}(\bar{\rho}).
\label{eqB15}
\end{equation}
Substitute Eq.~(\ref{eqB14}) into Eqs.~(\ref{eqB12}) and (\ref{eqB13}),
and use relations (\ref{eqB4})--(\ref{eqB6}) and (\ref{eqB15}).
After all those operations have been carried out,
the system of equations (\ref{eqB12}) takes its ultimate form,
\begin{eqnarray}
{\psi_{K^{\prime}n^{\prime}}(\rho)}&=&\frac{\pi{m}}{\hbar^{\,2}\rho^{2}}
\int\limits_{0}^{\infty}d{\bar{\rho}}\,{\bar{\rho}}^{\,3}\,P_{\pm}(\rho,\bar{\rho})
\int d{\Omega}\,u_{K^{\prime}n^{\prime}}^{\ast}\nonumber\\
&&\times\Bigl[U\sum\limits_{Kn}\psi_{Kn}(\bar\rho)u_{Kn}(\Omega)
+(V_{12}+V_{31})\Phi(\bar\rho,\Omega)\Bigr],
\label{eqB16}
\end{eqnarray}
\begin{equation}
P_{\pm}(\rho,\bar{\rho})=-\frac{2}{\pi}\int\limits_{0}^{\infty}dq\,q\,
\frac{J_{2}(q\rho)J_{2}(q\bar{\rho})}{q^{2}-k_{K^{\prime}}^{2}\mp{i0}}\,,
\label{eqB17}
\end{equation}
where $k_{K^{\prime}}^{2}\equiv{k_{K^{\prime}}^{2}(\rho)}=
k_{0}^{2}-K^{\prime}(K^{\prime}+4)/{\rho^{2}}$,
$k_{0}^{2}=2mE/\hbar^{2}$, and $E$ is the total energy of the system.
An integral of type (\ref{eqB17}) can be calculated analytically~\cite{39},
\begin{equation*}
\int{dx}\,x\,\frac{J_{\nu}(ax)J_{\nu}(bx)}{x^{2}+c^{2}}=
\end{equation*}
\begin{equation}
\left\{{
\begin{array}{*{20}c}
I_{\nu}(bc)K_{\nu}(ac),\, 0\!<\!b\!<\!a,\,\text{Re}{\,c}\!>\!0,\,
\text{Re}{\,\nu}\!>\!-1; \\
I_{\nu}(ac)K_{\nu}(bc),\, 0\!<\!a\!<\!b,\,\text{Re}{\,c}\!>\!0,\,
\text{Re}{\,\nu}\!>\!-1; \\
I_{\nu}(-bc)K_{\nu}(-ac),\, 0\!<\!b\!<\!a,\,\text{Re}{\,c}\!<\!0,\,
\text{Re}{\,\nu}\!>\!-1; \\
I_{\nu}(-ac)K_{\nu}(-bc),\, 0\!<\!a\!<\!b,\,\text{Re}{\,c}\!<\!0,\,
\text{Re}{\,\nu}\!>\!-1. \\
\end{array}}\right.
\label{eqB18}
\end{equation}
Now, introducing functions (\ref{eq10})--(\ref{eq13}) and taking
Eqs.~(\ref{eqB17}) and (\ref{eqB18}) into account, the system of
equations (\ref{eqB16}) can be rewritten in compact form (\ref{eq7}).

\setcounter{equation}{0}
\section{\normalsize{Spin-isospin functions for $^3$\!H and $^3$\!He nuclei}}
\hspace{\parindent}
Omitting brackets in the notation for symmetry states (\ref{eq15}),
$\xi^{\lbrack g]}\equiv{\xi^{g}}$, let us present the spin-isospin functions
of three-nucleon system at $S\!=\!1/2$ and $T\!=\!1/2$ as linear
combinations of spin, $\chi$, and isospin, $\zeta$, component products~\cite{40}:
\begin{equation}
\xi^{s}\!=\!\frac{1}{\sqrt{2}}(\chi{^{\prime}}\zeta{^{\prime}}+
\chi{^{\prime\prime}}\zeta{^{\prime\prime}}),
\quad\xi^{a}\!=\!\frac{1}{\sqrt{2}}
(\chi{^{\prime}}\zeta{^{\prime\prime}}-
\chi{^{\prime\prime}}\zeta{^{\prime}}),
\label{eqC1}
\end{equation}
\begin{equation}
\xi{^{\prime}}\!=\!\frac{1}{\sqrt{2}}(\chi{^{\prime}}\zeta{^{\prime\prime}}+
\chi{^{\prime\prime}}\zeta{^{\prime}}),\quad\xi{^{\prime\prime}}\!=
\!\frac{1}{\sqrt{2}}(\chi{^{\prime}}\zeta{^{\prime}}-\chi{^{\prime\prime}}
\zeta{^{\prime\prime}}).
\label{eqC2}
\end{equation}
Here, $\xi^{s}$ and $\xi^{a}$ are the functions completely symmetric and
completely antisymmetric, respectively, with respect to the permutation
of any pair of nucleons; they are basis functions for two corresponding
one-dimensional representations of permutation group for three nucleons.
Besides, $\xi{^{\prime}}$ and $\xi{^{\prime\prime}}$ are the basis
functions for a two-dimensional irreducible representation of the same group,
which are characterized by intermediate (mixed) symmetry.

The spin and isospin wave functions look like
\begin{equation}
\chi{^{\prime}}\!=\!\sqrt{2/3}T{^{\prime}}(\alpha_{2}\alpha_{3}\beta_{1}),\quad
\chi{^{\prime\prime}}\!=\!\sqrt{2/3}T{^{\prime\prime}}(\alpha_{2}\alpha_{3}\beta_{1});
\label{eqC3}
\end{equation}
\begin{equation}
^{3}{\mbox{H}}:\,\,\zeta{^{\prime}}\!=\!\sqrt{2/3}T{^{\prime}}(b_{2}b_{3}a_{1}),\quad
\zeta{^{\prime\prime}}\!=\!\sqrt{2/3}T{^{\prime\prime}}(b_{2}b_{3}a_{1}),
\label{eqC4}
\end{equation}
\begin{equation}
^{3}{\mbox{He}}:\,\,\zeta{^{\prime}}\!=\!\sqrt{2/3}T{^{\prime}}(a_{2}a_{3}b_{1}),\quad
\zeta{^{\prime\prime}}\!=\!\sqrt{2/3}T{^{\prime\prime}}(a_{2}a_{3}b_{1}),
\label{eqC5}
\end{equation}
where
\begin{equation}
T{^{\prime}}\!=\!(\!\sqrt{3}/2)[(13)-(12)],\quad T{^{\prime\prime}%
}\!=\!-(23)+[(13)+(12)]/2 \label{T12_def}%
\end{equation}
are permutation operators (the notation $(ij)$ stands for the permutation of
the $i$-th and $j$-th nucleons). The one-particle spin (isospin) wave
functions $\alpha_{j}$ and $\beta_{j}$ ($a_{j}$ and $b_{j}$) correspond to the
positive and negative, respectively, spin (isospin) projection of the $j$-th particle.

\end{document}